\documentclass[english,journal]{IEEEtran}
\usepackage[T1]{fontenc}
\usepackage[latin9]{inputenc}
\usepackage{textcomp}
\usepackage{mathrsfs}
\usepackage{amsmath}
\usepackage{amssymb}
\usepackage{graphicx}
\usepackage{esint}
\usepackage{relsize}

\makeatletter
\usepackage{cite}
\renewcommand{\fnum@figure}{Fig.~\thefigure}
\usepackage{flushend}

\usepackage{babel}
\begin{document}

\title{Models, Statistics, and Rates of\\ Binary Correlated Sources}

\author{Marco Martal\`o and Riccardo Raheli
\thanks{This paper was orally presented in part at the Information Theory and Applications (ITA) Workshop, San Diego, CA, USA,  February 2015. Marco Martal\`o is with the E-Campus University, Novedrate (CO), Italy, and the University of Parma, Italy. E-mail: marco.martalo@uniecampus.it. Riccardo Raheli is with the University of Parma, Italy. E-mail: raheli@unipr.it.}}

\maketitle
\begin{abstract}
This paper discusses and analyzes various models of binary correlated sources, which may be relevant in several distributed communication scenarios. These models are statistically characterized in terms of joint Probability Mass Function (PMF) and covariance. Closed-form expressions for the joint entropy of the sources are also presented. The asymptotic entropy rate for very large number of sources is shown to converge to a common limit for all the considered models. This fact generalizes recent results on the information-theoretic performance limit of communication schemes which exploit the correlation among sources at the receiver.
\end{abstract}


\section{Introduction\label{sec:Introduction}}
The efficient transmission of correlated signals, observed at various nodes, to one or more collectors is of wide interest in various scenarios, such as sensor networks~\cite{AkSuSaCa02}, and has been the subject of recent research attention. For instance,~\cite{JiPs05} discusses the spatial dependence between data according to the distribution of the nodes in the monitored area through empirical measurements. The design of efficient transmission schemes for correlated sources through orthogonal additive white Gaussian noise (AWGN) channels is discussed in~\cite{BaSe06}. In this case, the separation between source and channel coding is optimal and ultimate performance can be achieved by means of distributed source coding (DSC) followed by independent capacity-achieving channel coding~\cite{ShVe95,BaSe06}. An alternative solution is based on the use of joint source channel coding (JSCC) schemes, where proper codes are used to encode the correlated sources. In both cases, knowledge of the statistical source 
correlation is exploited at the joint decoder, whereas source encoding is performed separately~\cite{GaZh01}.

The design of universal codes for transmission of correlated sources, i.e., capacity-achieving codes for all possible channel parameters, is a current topic. Universal codes based on spatial coupling have been recently proposed~\cite{YePfNa12}. Orthogonal multiple access schemes with an arbitrary number of correlated sources have been recently addressed in~\cite{AbFeMaFrRa14}, where the asymptotic achievable region for increasing number of sources has been characterized in terms of individual channel capacities, for a specific correlation model, and pragmatic joint source-channel coded schemes have been proposed.

In this paper, we discuss various correlation models for an arbitrary number of binary sources, which may be of interest in several realistic communication scenarios. With the exception of~\cite{AbFeMaFrRa14}, there are not many papers in the literature discussing correlation models for a possibly large numbers of sources. In~\cite{LiRa12}, the authors proposed a correlation model based on a set of linear equations in the binary field. This model is shown to be related, under special conditions, to one of the binary symmetric channels (BSC)-based models discussed in this paper. Main contributions of this paper are the statistical characterization of such models in terms of joint probability mass function (PMF), covariance, and joint entropy of the sources. Moreover, considering the information sequence as a stochastic process in the spatial domain, we derive the asymptotic entropy rate for large number of sources and show it coincides for all the considered models.  Therefore, the asymptotic achievable 
region discussed in~\cite{AbFeMaFrRa14}, characterized in terms of the source entropy rate, can be inferred to be independent of the specific correlation model and pragmatic joint source-channel coded schemes may be expected to have similar asymptotic behavior regardless of the model.

This paper is structured as follows. In Section~\ref{sec:models}, we present various binary source correlation models. In Section~\ref{sec:pmf}, we statistically characterize these models, by deriving the joint PMF of their output sequences and the corresponding covariance matrices. In Section~\ref{sec:achievable}, we statistically characterize these schemes in terms of their joint entropy. In Section~\ref{sec:achievable2}, we use the joint entropy rate of these models to characterize the performance limit of orthogonal multiple access schemes transmitting correlated symbols. Finally, concluding remarks are given in Section~\ref{sec:Conclusions}.

\section{Source Correlation Models\label{sec:models}}
\begin{figure}
\begin{center}
\begin{tabular}{cc}
\includegraphics[width=0.4\textwidth]{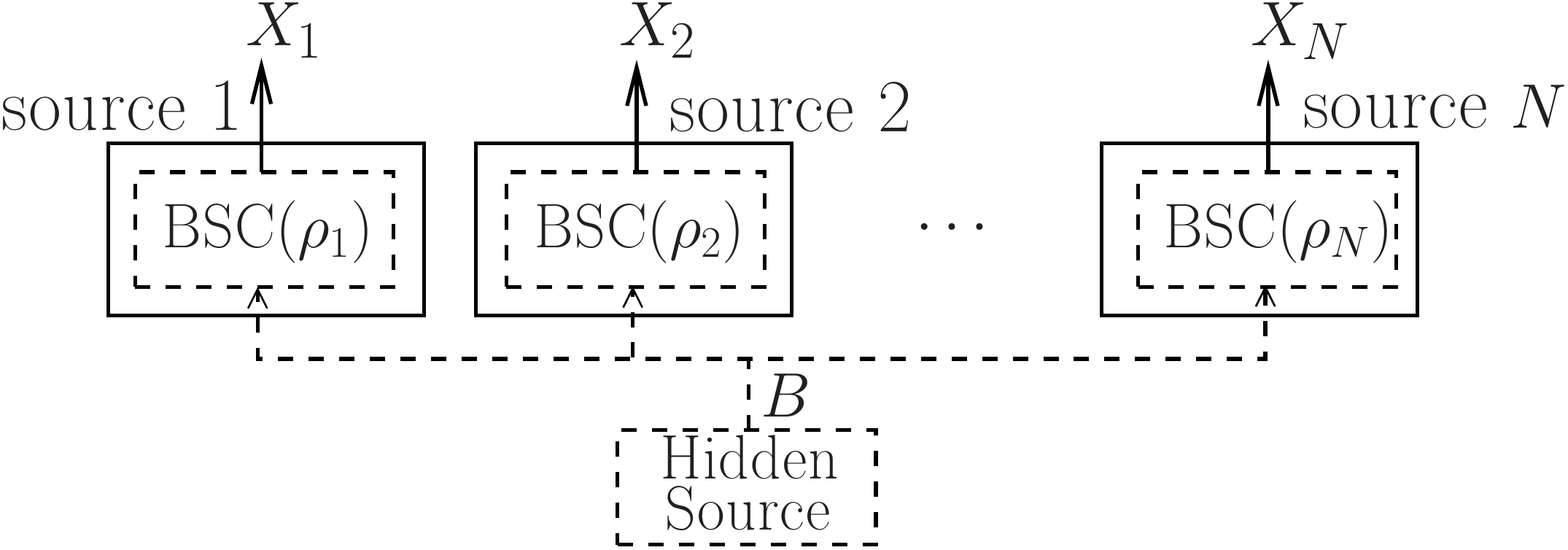}\\
(a)\\
\includegraphics[width=0.46\textwidth]{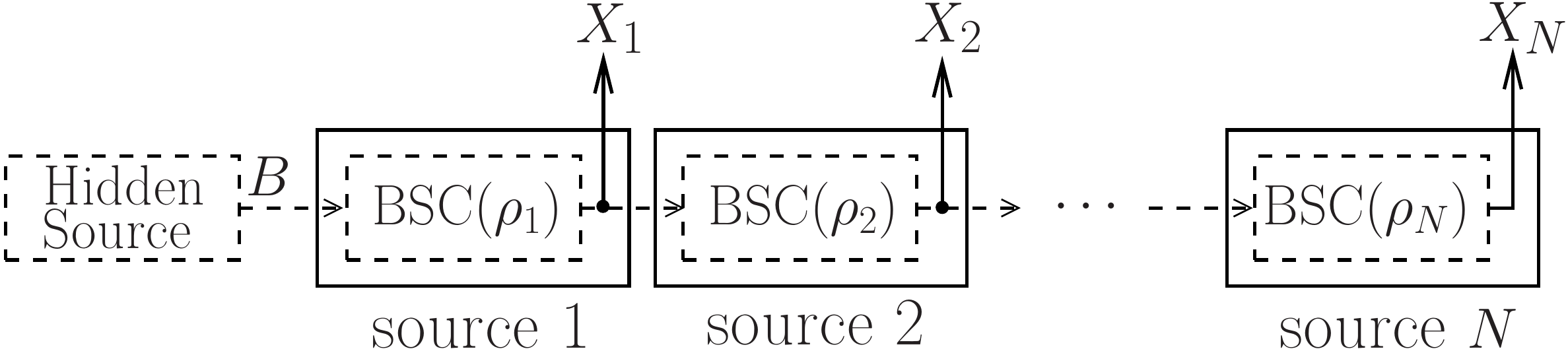}\\
(b)\\
\includegraphics[width=0.48\textwidth]{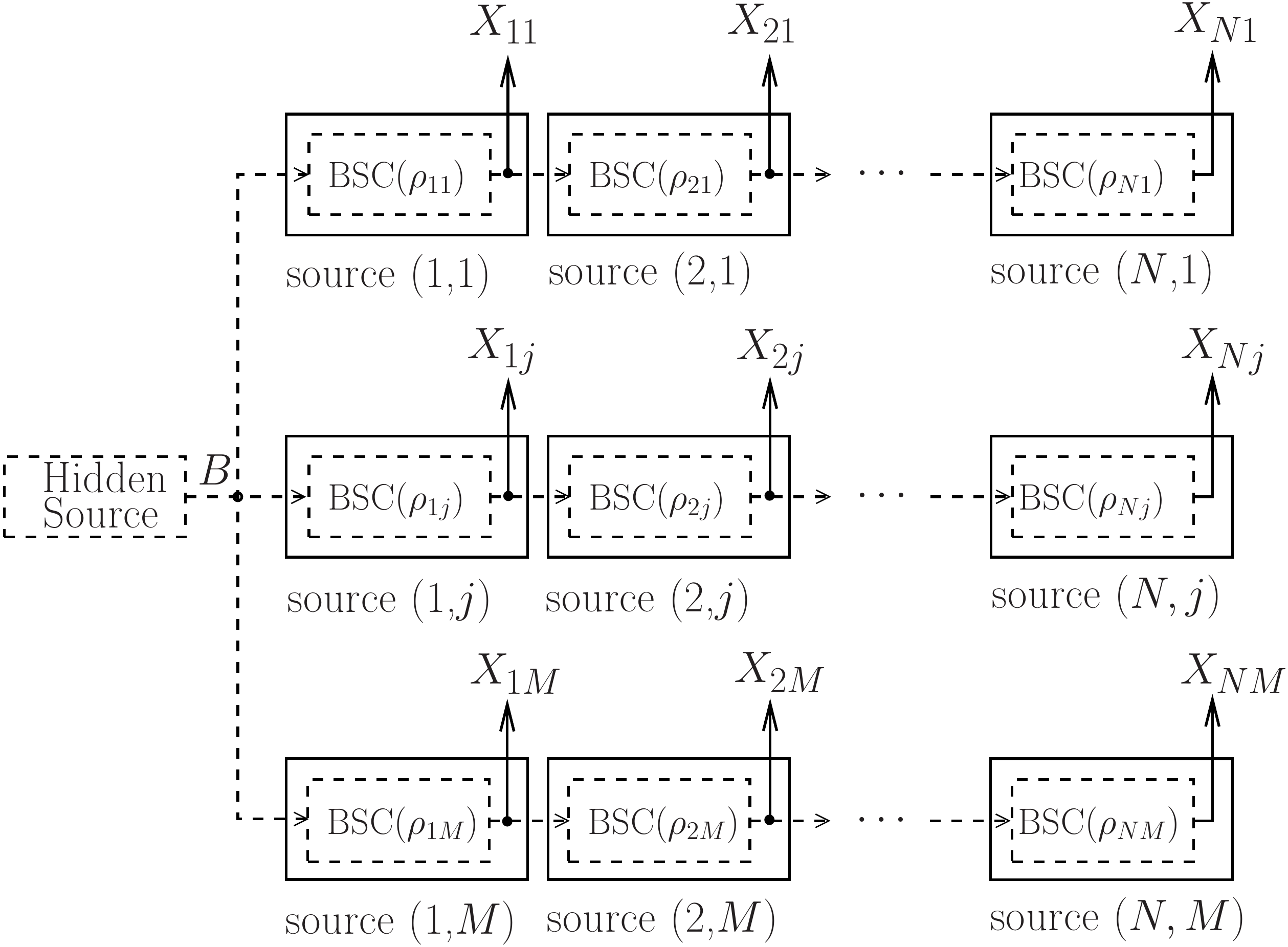}\\
(c)
\end{tabular}
\end{center}
\caption{Considered correlation models: (a) parallel, (b) serial, and (c) mixed.}
\label{img:correlation}
\end{figure}

Consider $N$ source nodes, possibly spatially distributed, which output (emit) binary information sequences $\pmb{X} = (X_1,X_2,\ldots,X_N)^T$, where $(\cdot)^T$ is the transpose operation. The binary information symbols are assumed to be marginally equiprobable, but correlated with each other according to a given PMF $P_{\pmb{X}}(\pmb{x})$, in which the $N$-element vector $\pmb{x}$ describes a possible realization of $\pmb{X}$. This scenario may be representative of a sensor network in which the sensors observe $N$ correlated physical quantities of interest. In Fig.~\ref{img:correlation}, possible correlation models are shown: (a) parallel, (b) serial, and (c) mixed.
%%%%%%%%%
%%%%%%%%%
In the parallel model~(a), the source symbols are the output of a set of parallel BSCs, with cross-over probability $1-\rho_\ell$, for $\ell=1,2,\ldots,N$, denoted as ${\rm BSC}(\rho_\ell)$, whose input is a hidden common information bit $B$. The $\ell$-th source symbol is given by
\begin{equation} \label{eq:corr_model}
X_\ell = B \oplus Z_\ell
\end{equation}
where $B$ is an independent equiprobable binary random variable, $\{Z_\ell\}$ are independent binary random variables with $P(Z_\ell=0)=\rho_\ell$, $1/2\leq\rho_\ell\leq1$ for $\ell=1,2,\ldots,N$, and $\oplus$ denotes a modulo-2 sum. Obviously, if $\rho_\ell = 0.5$ there is no correlation among the binary information symbols $\{X_\ell\}_{\ell=1}^N$, whereas if $\rho_\ell = 1$ they are identical with probability 1.

In Fig.~\ref{img:correlation}~(b), a possible serial correlation model is shown, in which the source symbols are correlated by a cascade of BSCs\footnote{The first BSC does not play any role in the serial model and could be omitted---it is kept for notational consistency with the other models.} with cross-over probability $1-\rho_\ell$. Again the output symbols are uncorrelated for $\rho_\ell=0.5$, whereas they are equal with probability 1 for $\rho_\ell=1$. 

A more general ``mixed'' case with a number $m$ of serial chains is shown in Fig.~\ref{img:correlation}~(c), in which $1-\rho_{ij}$ denotes the cross-over probability of the $i$-th BSC on the $j$-th chain. In this case, the correlated data at the $\ell$-th, $\ell=1,2,\ldots,N$, source on the $j$-th branch, $j=1,2,\ldots,M$, can be expressed as
$$
X_{\ell j} = B \oplus \underbrace{\sum_{i=1}^\ell \!\!\!\!\!\oplus \hspace*{2mm} Z_{ij}}_{\triangleq Z'_{\ell j}} = B \oplus Z'_{\ell j}
$$
where the symbol $\displaystyle\sum\!\!\!\!\!\oplus$ denotes modulo-2 sums. The random variable $Z'_{\ell j}$ can be easily characterized by its distribution~\cite[Lemma~4.1]{Ga63}
\begin{equation} \label{eq:prob_Gallager}
p_{\ell j} \triangleq P(Z'_{\ell j} = 0) = \frac{1}{2} \left[ 1 + \mathlarger{\prod}\limits_{i=1}^{\ell}(2\rho_{ij}-1)\right] .
\end{equation}
Note that for $M=1$, this mixed model reduces to the serial one of Fig.~\ref{img:correlation}~(b) and the index $j$ in (\ref{eq:prob_Gallager}) can be dropped.

A model based on a set of binary linear equations was considered in~\cite{LiRa12}:
\begin{equation} \label{eq:corr_model2}
\pmb{A}\pmb{X} = \pmb{Z}
\end{equation}
where $\pmb{A}$ is a binary matrix (whose entries are equal to either 0 or 1) defining the set of equations and $\pmb{Z}=(Z_1,Z_2,\ldots,Z_N)^T$ is a vector of independent Bernoulli binary random variables with parameter $\rho_\ell=P(Z_\ell=0)$. Note that matrix operations are performed in the binary field. As a special case, (\ref{eq:corr_model2}) may encompass a set of recursive equations of the form
\begin{equation} \label{eq:literature_model}
\sum_{i=0}^{\min\{D,\ell-1\}}\!\!\!\!\!\!\!\!\!\!\!\!\!\!\!\oplus \hspace*{2mm} A_{\ell,\ell-i} X_{\ell-i} = Z_\ell
\end{equation}
in which $D$ is the recursion depth, $A_{\ell \ell}=1$, $A_{\ell k}=0$ or 1 for $k=\ell-\min\{D,\ell-1\},\ldots,\ell-1$, and $A_{\ell k}=0$ for $k>\ell$. Since we are considering binary random variables, (\ref{eq:literature_model}) can be also rewritten as
\begin{equation} \label{eq:literature_model2}
X_\ell = \left\{
\begin{array}{ll}
Z_\ell & \ell=1\\
Z_\ell\, \, \, \oplus \!\!\mathlarger{\sum}\limits_{i=1}^{\min\{D,\ell-1\}}\!\!\!\!\!\!\!\!\!\!\!\!\!\!\!\oplus \hspace*{2mm} A_{\ell,\ell-i} X_{\ell-i} & \ell=2,3,\ldots,N .
\end{array}
\right.
\end{equation}
The model in (\ref{eq:literature_model2}) describes a recursive binary filter with input $Z_\ell$ and output $X_\ell$. In the special case of $D=1$, $A_{\ell,\ell-1}=1$, and $\rho_1=1/2$, the model in (\ref{eq:literature_model2}) is equivalent to the serial one in Fig.~\ref{img:correlation}~(b).

If the matrix $\pmb{A}$ is invertible in the binary field, then
\begin{equation} \label{eq:inverse}
\pmb{X} = \pmb{A}^{-1}\pmb{Z}
\end{equation}
and, therefore, there exists a one-to-one correspondence between $\pmb{X}$ and $\pmb{Z}$. A case where the inverse exists is the recursive model in (\ref{eq:literature_model}), as it is shown in Appendix~\ref{app:invertible}, where a few cases of interest are also analyzed. For simplicity, in the rest of the paper we will assume that the matrix $\pmb{A}$ is invertible. Note that, invertibility may lead to uncorrelated binary sources in the special case of matrices with constant row weight equal to $d_{\rm c}$, as shown in~\cite{ToRoGu11}.

\section{Statistical Characterization\label{sec:pmf}}
According to the parallel correlation model (\ref{eq:corr_model}) in Fig.~\ref{img:correlation}~(a), the joint PMF of the information symbols at the output of the $N$ nodes can be computed. By straightforward manipulations, one can show that
\begin{eqnarray}
P_{\pmb{X}}(\pmb{x}) &=& \sum_{b=0,1} P_{\pmb{X}}(\pmb{x}|B=b)P_B(b) \nonumber \\
&& \hspace*{-15mm}=\frac{1}{2}\left[\prod_{\ell\in\mathcal{S}_0}\rho_\ell\prod_{k\in\mathcal{S}_1}(1-\rho_k)+\prod_{\ell\in\mathcal{S}_0}(1-\rho_\ell)\prod_{k\in\mathcal{S}_1}\rho_k\right]
 \label{eq0AA}
\end{eqnarray}
where $\mathcal{S}_0$ and $\mathcal{S}_1$ is a partition of the set $\{1,2,\ldots,N\}$ specifying the positions of zeros and ones in $\pmb{x}$, respectively. In the special case of $\rho_\ell=\rho$, $\ell=1,2,\ldots,N$, one obtains
\begin{equation} \label{eq0AA1}
P_{\pmb{X}}(\pmb{x})=\frac{1}{2}\left[\rho^{n_{\rm z}} (1-\rho)^{N-n_{\rm z}} + (1-\rho)^{n_{\rm z}} \rho^{N-n_{\rm z}}\right]
\end{equation}
where $n_{\rm z}$ is the number of zeros in $\pmb{x}$.

In the serial case of Fig.~\ref{img:correlation}~(b), using the chain rule, one has
\begin{eqnarray}
P_{\pmb{X}}(\pmb{x}) &=& \sum_{b=0,1} P_{\pmb{X}}(\pmb{x}|B=b)P_B(b) \nonumber \\
&& \hspace*{-20mm}=\sum_{b=0,1} P_B(b) P_{X_1}(x_1|B=b) \prod_{\ell=2}^N P_{X_\ell}(x_\ell|X_{\ell-1}=x_{\ell-1})  \label{eq:serial_general}
\end{eqnarray}
where we have used the fact that, given $X_{\ell-1}$, $X_\ell$ is conditionally independent from the previous source symbols $X_{\ell-2},\ldots,X_1$ and $B$. After further simple manipulations, one can write
$$
P_{\pmb{X}}(\pmb{x}) = \frac{1}{2}\prod_{\ell\in\mathcal{L}_0}\rho_\ell\prod_{k\in\mathcal{L}_1}(1-\rho_k)
$$
where $\mathcal{L}_0$ and $\mathcal{L}_1$ is a partition of the set $\{2,3,\ldots,N\}$ specifying the positions where $x_\ell=x_{\ell-1}$ and $x_\ell\neq x_{\ell-1}$, respectively. In the special case of $\rho_\ell=\rho$, $\ell=2,3,\ldots,N$, denoting as $n'_{\rm z}$ the cardinality of the set $\mathcal{L}_0$, one obtains
$$
P_{\pmb{X}}(\pmb{x}) = \frac{1}{2} \rho^{n'_{\rm z}} (1-\rho)^{N-n'_{\rm z}-1} .
$$

\begin{figure*}
\centering
\begin{tabular}{ccc}
\includegraphics[width=0.31\textwidth]{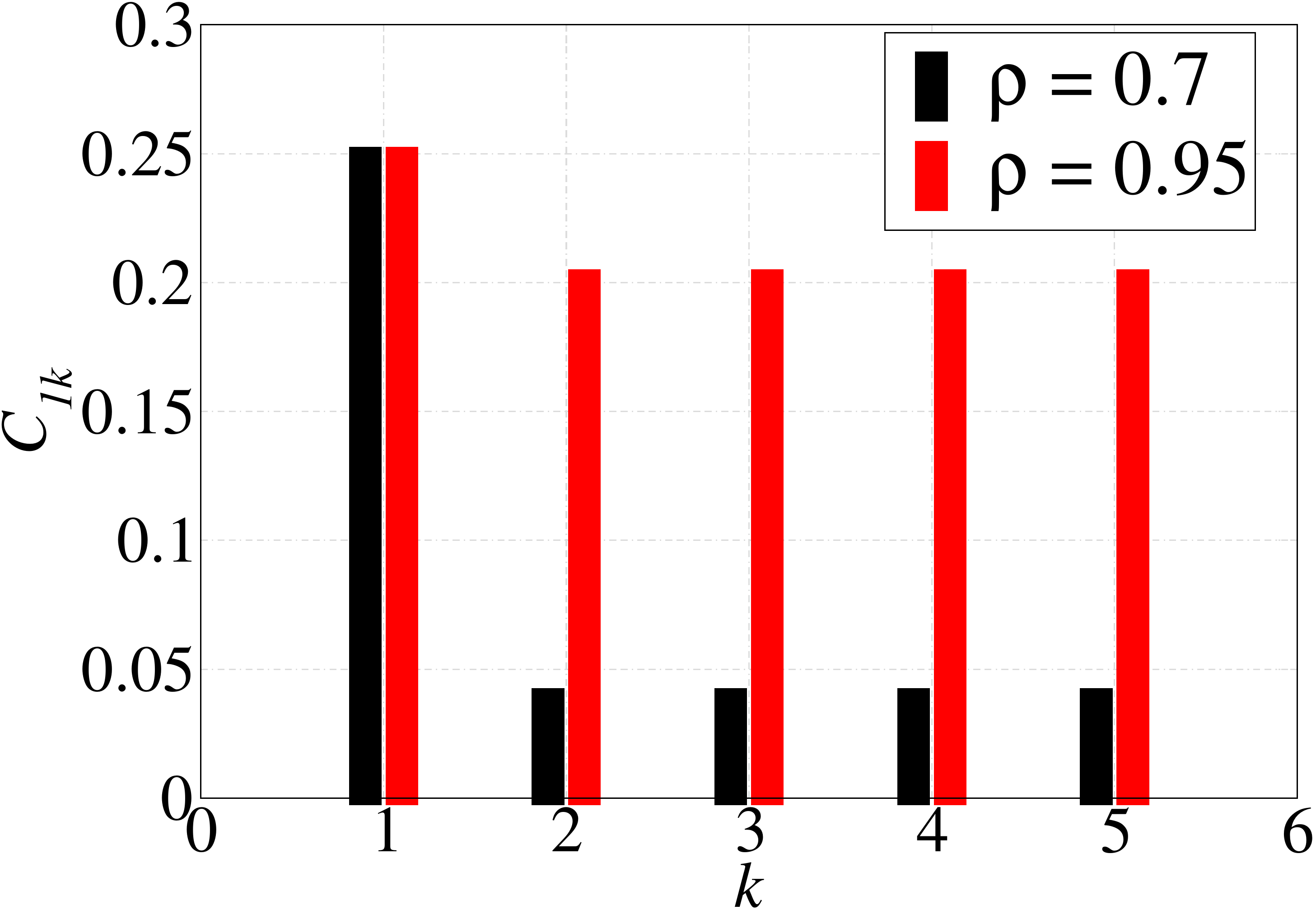} & \includegraphics[width=0.31\textwidth]{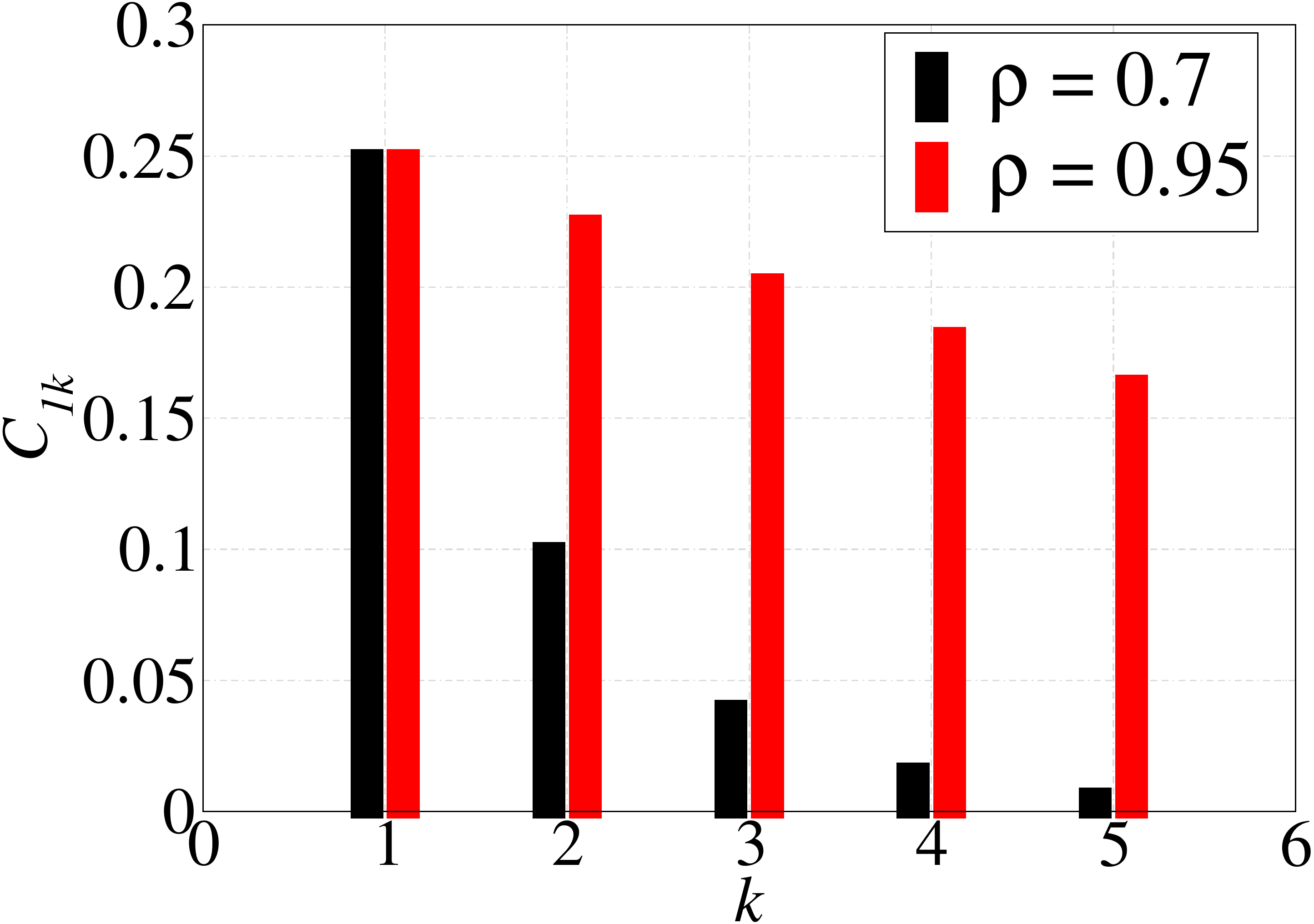} & \includegraphics[width=0.31\textwidth]{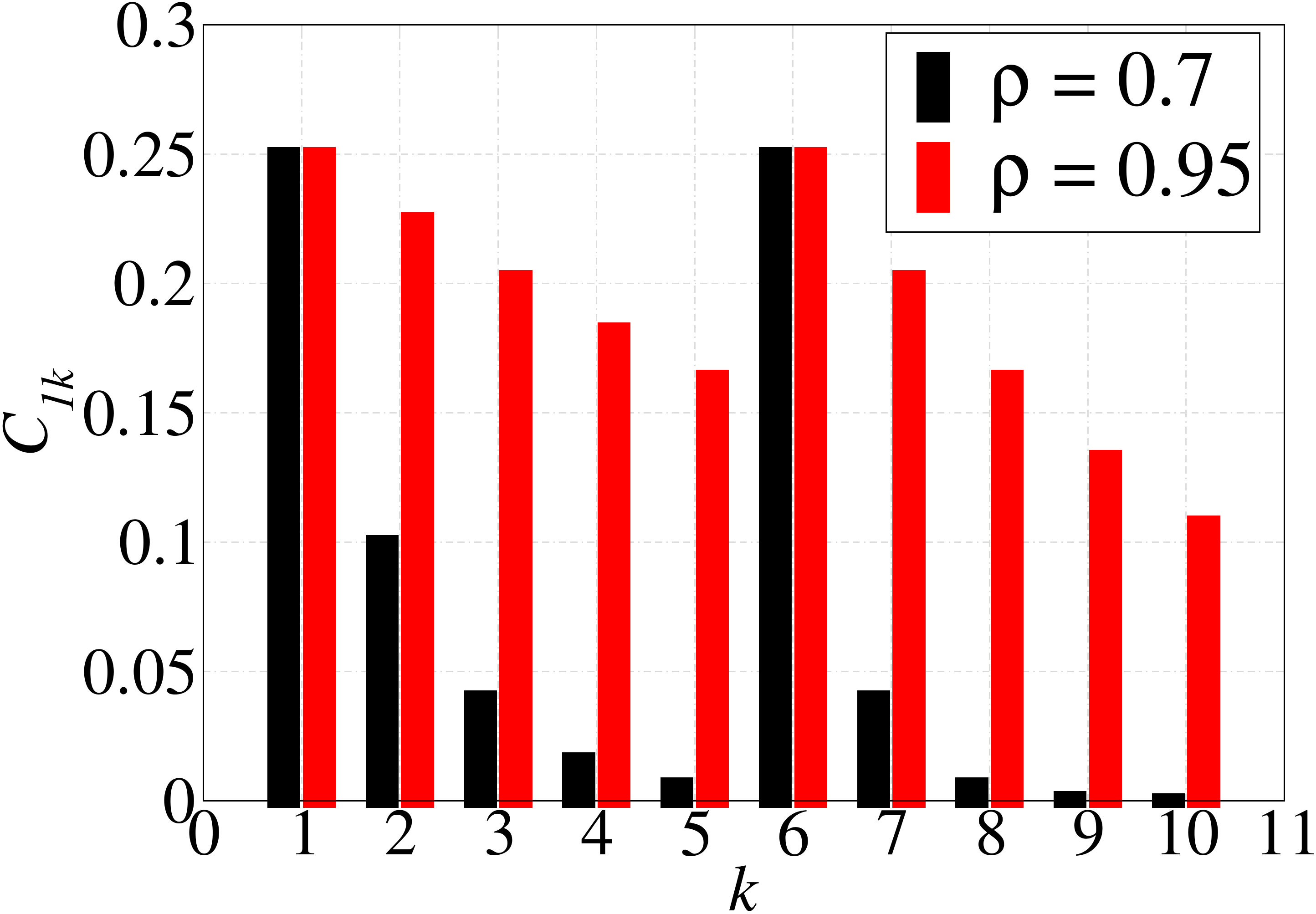}\\
(a) & (b) & (c)
\end{tabular}
\caption{Coefficients of the first row of the covariance matrix $\pmb{C}_X$ for $N=5$ sources and three models: (a) parallel, (b) serial, and (c) mixed with $M=2$. Two values of $\rho$ are considered: 0.7 (left bars in each figure) and 0.95 (right bars in each figure).}
\label{img:histo}
\end{figure*}

In the general mixed case of Fig.~\ref{img:correlation}~(c), let us denote as
$$
X_{1j}^N \triangleq \left(X_{1j},\ldots,X_{Nj}\right) \hspace*{10mm} j=1,2,\ldots,M
$$
the set of the $N$ source symbols on the $j$-th branch. One can generalize the result of the serial case by writing:
\begin{eqnarray*}
P_{\pmb{X}}(\pmb{x}) &=& \sum_{b=0,1} P_{\pmb{X}}(\pmb{x}|B=b)P_B(b) \\ 
&=& \sum_{b=0,1} \prod_{j=1}^M P_{X_{1j}^N}(x_{1j}^N|B=b)P_B(b)\\
&=& \sum_{b=0,1} \prod_{j=1}^M P_B(b) P_{X_{1j}}(x_{1j}|B=b)\\
&&\hspace*{8mm}\cdot\prod_{\ell=2}^N P_{X_{\ell j}}(x_{\ell j}|X_{\ell-1,j}=x_{\ell-1,j})
\end{eqnarray*}
where, in the first line, we have used the fact that, conditionally on $B$, the branches are independent, and the second line is equivalent to (\ref{eq:serial_general}) applied to each branch.

Finally, the PMF for correlated sources following the model in~\cite{LiRa12} can be characterized, by resorting to (\ref{eq:corr_model2}), as
\begin{equation} \label{eq:identity}
P_{\pmb{X}}(\pmb{x}) = P_{\pmb{Z}}(\pmb{A}\pmb{x}) .
\end{equation}
Using simple manipulations, we have
$$
P_{\pmb{X}}(\pmb{x}) = \prod_{\ell\in\mathcal{S}'_0}\rho_\ell\prod_{k\in\mathcal{S}'_1}(1-\rho_k)
$$
where $\mathcal{S}'_0$ and $\mathcal{S}'_1$ is a partition of the set $\{1,2,\ldots,N\}$ specifying the positions of zeros and ones in $\pmb{A}\pmb{x}$, respectively.

The covariance matrix for the considered correlation models is defined as
$$
\pmb{C}_X = \pmb{R}_X - \pmb{\mu}_X\pmb{\mu}_X^T .
$$
For the parallel and serial models, the elements of the correlation matrix $\pmb{R}_X$ and the mean vector $\pmb{\mu}_X$ are, respectively,
$$
\begin{array}{lll}
R_{ik} & = & \mathbb{E}[X_i X_k]\\
\mu_i & = & \mathbb{E}[X_i]
\end{array}
\hspace*{10mm}
i,k=1,2,\ldots,N.
$$
Note that $R_{ik}=\mathbb{E}[X_i X_k]=P(X_i=X_k=1)$, since the data are binary. The elements of the vector $\pmb{\mu}_X$ can be obtained as
$$
\mu_i = \dfrac{1}{2} .
$$
$\pmb{R}_X$ depends on the considered correlation model, but in all cases
$$
R_{ii}=\mathbb{E}\left[X_i^2\right]= 1\cdot P(X_i=1) = \frac{1}{2}
$$
since, for all considered models, $P(X_i=1)=0.5$.

In the parallel case of Fig.~\ref{img:correlation}~(a), for $i,k=1,2,\ldots,N$ ($i\neq k$) one obtains
\begin{eqnarray}
\mathbb{E}[X_i X_k] &=& \frac{1}{2} \sum_{b=0,1} P(X_i=X_k=1|B=b) \nonumber \\
&=& \frac{1}{2}\left[\rho_i\rho_k + (1-\rho_i)(1-\rho_k)\right] \nonumber \\
&=& \frac{1}{2}\left[1 - (\rho_i+\rho_k) + 2\rho_i\rho_k\right] \label{eq:corr1}
\end{eqnarray}
which reduces, for $\rho_\ell=\rho$, $\ell=1,2,\ldots,N$, to
$$
\mathbb{E}[X_i X_k] = \frac{1}{2}\left[1-2\rho+2\rho^2\right] .
$$
As expected, the correlation is independent of the indices of the considered sources. Note that if $\rho=1/2$, $C_{ik}=1/4$ for $i=k$ and zero otherwise, i.e., data are uncorrelated.

For the serial correlation model in Fig.~\ref{img:correlation}~(b) and $i\neq k$, one can write
\begin{eqnarray*}
P(X_i=X_k=1) &=& P(X_k=1|X_i=1) P(X_i=1) \\
&=& \dfrac{1}{2}P(X_k=1|X_i=1)
\end{eqnarray*}
where the conditional probability $P(X_k=1|X_i=1)$ can be computed noting that $X_i=X_k=1$ if the BSCs of indices $i+1,i+2,\ldots,k$ flip an even number of times. Therefore, by arguments similar to those in~~\cite[Lemma~4.1]{Ga63} we obtain
$$
P(X_k=1|X_i=1) = \frac{1}{2}\left[1+\mathlarger{\prod}\limits_{\ell=i+1}^k (2\rho_\ell-1)\right]
$$
and, therefore,
\begin{equation} \label{eq:corr2}
\mathbb{E}[X_i X_k] = \frac{1}{4}\left[1+\mathlarger{\prod}\limits_{\ell=i+1}^k (2\rho_\ell-1)\right] .
\end{equation}
For the special case of $\rho_\ell=\rho$, $\ell=2,3,\ldots,N$, denoting the number of hops in the chain of BSCs as $L=|i-k|$, (\ref{eq:corr2}) reduces to
$$
\mathbb{E}[X_i X_k] = \frac{1}{4}\left[1+(2\rho-1)^L\right] .
$$
Note that if $\rho=1/2$, $C_{ik}=1/4$ for $i=k$ and zero otherwise, i.e., data are uncorrelated.

In the mixed scenario of Fig.~\ref{img:correlation}~(c), the covariance matrix has size $NM\times NM$. The element $C_{ik}$, for $i,k=1,2,\ldots,NM$, can be defined as the covariance between the source symbols $X_{\ell_1 m_1}$ and $X_{\ell_2 m_2}$, where
$$
i = (m_1-1)N + \ell_1 \hspace*{10mm} k = (m_2-1)N + \ell_2 .
$$
In particular, the following two cases may occur:
\begin{itemize}
\item if $m_1=m_2$, i.e., $X_{\ell_1 m_1}$ and $X_{\ell_2 m_1}$ belong to the same branch, the result in (\ref{eq:corr2}) can be applied;
\item if $m_1\neq m_2$, i.e., $X_{\ell_1 m_1}$ and $X_{\ell_2 m_2}$ belong to different branches, the results for two sources in a parallel scheme can be applied replacing $\rho_i$ and $\rho_k$ in (\ref{eq:corr1}) with $p_{\ell_1 m_1}$ and $p_{\ell_2 m_2}$, respectively, as defined in (\ref{eq:prob_Gallager}).
\end{itemize}
Obviously, it is still verified that, if $\rho_{ik}=\rho=1/2$, $C_{ik}=1/4$ for $i=k$ and zero otherwise, i.e., data are uncorrelated.

Finally, the covariance matrix can be also considered for the correlation model (\ref{eq:corr_model2}) based on a set of binary linear equations. Using (\ref{eq:inverse}), one can write
\begin{eqnarray*}
\pmb{C}_X &=& \pmb{R}_X - \pmb{\mu}_X\pmb{\mu}_X^T \\
&=& \mathbb{E}\left[\pmb{A}^{-1}\pmb{Z}\pmb{Z}^T(\pmb{A}^{-1})^T\right] - \mathbb{E}\left[\pmb{A}^{-1}\pmb{Z}\right]\mathbb{E}\left[\pmb{Z}^T(\pmb{A}^{-1})^T\right]\\
&=& P\left(\pmb{A}^{-1}\pmb{Z}\pmb{Z}^T(\pmb{A}^{-1})^T=\pmb{J}_N\right) \\
&&-P\left(\pmb{A}^{-1}\pmb{Z}=\pmb{1}_N\right)P\left(\pmb{Z}^T(\pmb{A}^{-1})^T=\pmb{1}_N\right)
\end{eqnarray*}
where we have used the fact that the mean value of a binary random variable is equal to the probability that the variable is equal to 1, $\pmb{J}_N$ is the all-one matrix of size $N\times N$, and $\pmb{1}_N$ is the all-one column vector of length $N$. Note that a closed-form solution for $\pmb{C}_X$ is not readily available as it depends on the particular structure of $\pmb{A}$.

Fig.~\ref{img:histo} shows the coefficients of the first row of the covariance matrix $\pmb{C}_X$ for $N=5$ sources, $\rho_\ell=\rho_{ik}=\rho$, and the three models: (a) parallel, (b) serial, and (c) mixed with $M=2$. Two values of $\rho$ are considered: 0.7 (left bars in each figure) and 0.95 (right bars in each figure). Only the first row is considered, since for parallel and serial models with constant $\rho$ the covariance matrix is symmetric and Toeplitz. In the mixed case, instead, for constant $\rho$, $\pmb{C}_X$ has the following block structure
$$
\pmb{C}_X = \left[
\begin{array}{cccc}
\pmb{C}_1 & \pmb{C}_2 & \cdots & \pmb{C}_M\\
\pmb{C}_2^T & \pmb{C}_1 & \cdots & \pmb{C}_{M-1}\\
\vdots & \vdots & \vdots & \vdots \\
\pmb{C}_M^T & \pmb{C}_{M-1}^T & \cdots & \pmb{C}_1
\end{array}
\right]
$$
where $\pmb{C}_i$ is the covariance matrix, of size $N\times N$, between sources on chains with separation $i-1$. In other words, zero separation means that the sources are in the same chain, separation 1 means that sources are on adjacent chains, and so on. In this case as well, the first row is sufficient to characterize the entire matrix. In the figure, one can observe that the first coefficient is equal to the symbol variance ($1/4$) in all cases. Moreover, in the parallel model~(a), all coefficients $C_{1k}$ for $k$ from 2 to 5 are equal due to the fact that the pairwise probabilities are identical, regardless of the source index. Note also that the higher the value of $\rho$, the higher the covariance elements, since data are more and more correlated. In the serial model~(b), the covariance decreases with $k$, since a larger number of BSCs is present between the sources, which become more and more uncorrelated. In the mixed model~(c), recall that indices from $k=1$ to $k=5$ correspond to the first 
chain of BSCs, whereas indices from $k=6$ to $k=10$ correspond to the second chain. As expected, symbols in the second chain are less correlated with the first symbol $X_{11}$, than those in the first one.

\begin{figure}
\centering
\begin{tabular}{c}
\includegraphics[width=0.3\textwidth]{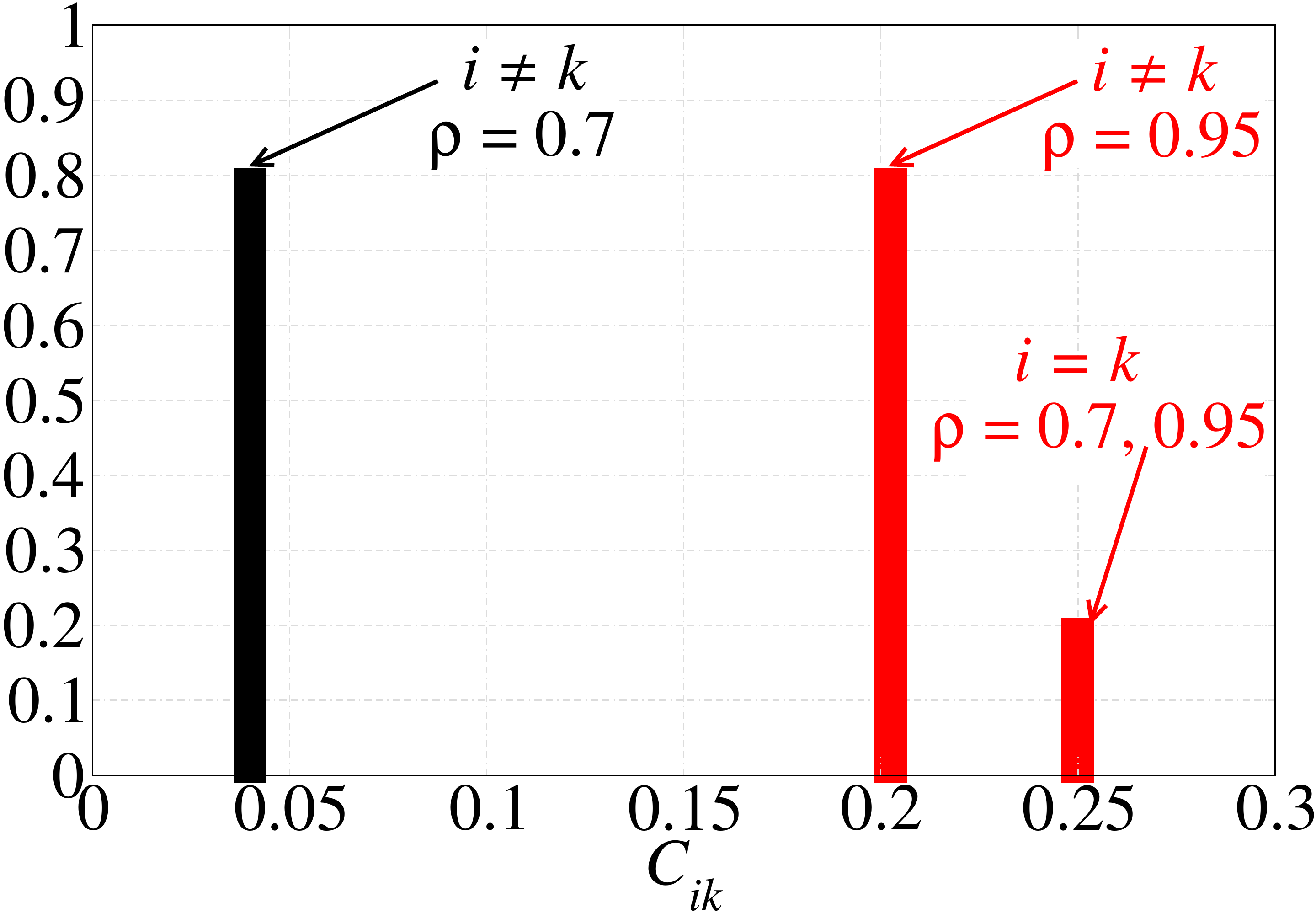} \\
(a)\\
\includegraphics[width=0.3\textwidth]{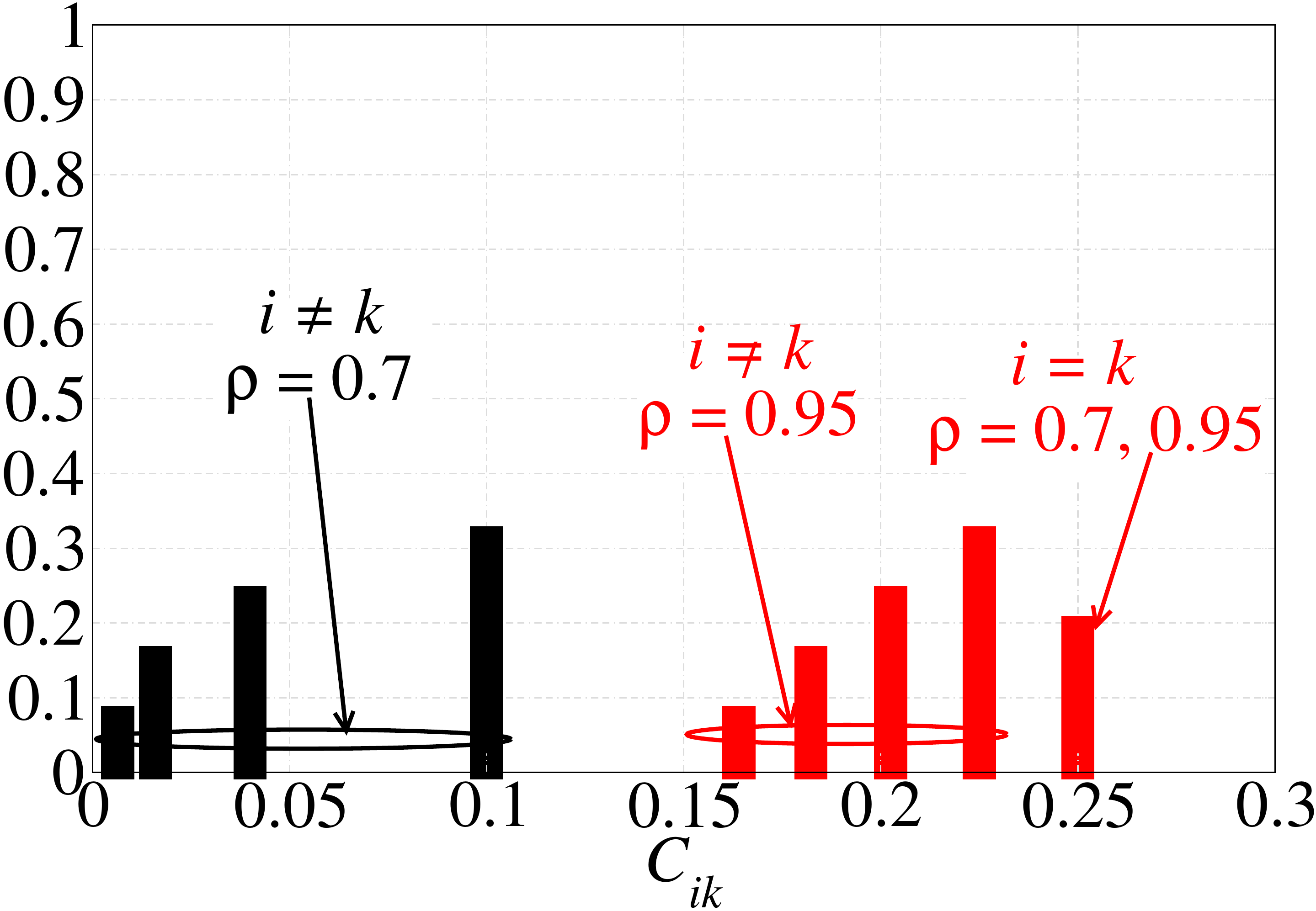}\\
(b)
\end{tabular}
\caption{Histogram of the values of the matrix $\pmb{C}_X$ for $N=5$ sources and two models: (a) parallel and (b) serial. Two values of $\rho$ are considered: 0.7 and 0.95.}
\label{img:covariance}
\end{figure}
An alternative way to view the covariance matrix is by the histogram of its values shown in Fig.~\ref{img:covariance} for $N=5$ sources and two models: (a) parallel and (b) serial. Two values of $\rho_\ell=\rho$ are considered: 0.7 and 0.95. One can observe that in the parallel model, only two values of $C_{ik}$ are allowed, since the pairwise probabilities are the same for any pair of sources. Moreover, for high correlation (e.g., $\rho=0.95$) larger values than those for small correlation (e.g., $\rho=0.7$) are obtained, which is in agreement with the fact that $C_{ik}\rightarrow0$ for $i\neq k$ if $\rho\rightarrow1/2$.

\section{Source Entropy Rate\label{sec:achievable}}
To compute the joint entropy of the $N$ sources for the considered correlation models, we denote it as $H(X_1^N)$, where the notation $X_a^b$ indicates the sequence $(X_a,X_{a+1},\ldots,X_b)$. Let us consider the joint entropy of $X_1^N$ and $B$:
\begin{equation} \label{eq:proof3}
H(B,X_1^N) = H(B) + H(X_1^N|B) .
\end{equation}
For the parallel model (\ref{eq:corr_model}) of Fig.~\ref{img:correlation}~(a), one has
$$
H(X_1^N|B) = \sum_{\ell=1}^N H_{\rm b}(\rho_\ell)
$$
where $H_{\rm b}(\rho_\ell)$ is the entropy of a binary random variable with parameter $\rho_\ell$. Since $B$ is a uniformly distributed binary random variable, $H(B)=H_{\rm b}(0.5)=1$. Therefore, one obtains:
$$
H(B,X_1^N) = 1 + \sum_{\ell=1}^N H_{\rm b}(\rho_\ell) .
$$
As in (\ref{eq:proof3}), it is also possible to write
$$
H(B,X_1^N) = H(X_1^N) + H(B|X_1^N) = 1+ \sum_{\ell=1}^N H_{\rm b}(\rho_\ell)
$$
and, therefore,
\begin{equation}
H(X_1^N) = 1 + \sum_{\ell=1}^N H_{\rm b}(\rho_\ell) - H(B|X_1^N) \label{eq:hn_th3} .
\end{equation}
By definition of entropy, the last term is non negative and the following upper bound (UB) is obtained
\begin{equation} \label{eq:a}
H(X_1^N) \leq 1+ \sum_{\ell=1}^N H_{\rm b}(\rho_\ell) .
\end{equation}
Moreover, since conditioning reduces entropy~\cite{CoTh06}, it also follows that
$$
H(B|X_1^N) \leq H(B|X_1) = H_{\rm b}(\rho_1) .
$$
Using this in (\ref{eq:hn_th3}), one obtains the lower bound (LB):
\begin{equation}
H(X_1^N) \geq 1 + \sum_{\ell=1}^N H_{\rm b}(\rho_\ell) - H_{\rm b}(\rho_1) = 1 + \sum_{\ell=2}^N H_{\rm b}(\rho_\ell) . \label{eq:b}
\end{equation}
Combining (\ref{eq:a}) and (\ref{eq:b}) and analyzing the limit for large number of sources, one obtains
\begin{equation} \label{eq:limit2}
\lim_{N\rightarrow+\infty}\frac{H(X_1^N)}{N} = \overline{H}_{\rm b}
\end{equation}
where
$$
\overline{H}_{\rm b} \triangleq \lim_{N\rightarrow+\infty}\frac{1}{N}\sum_{\ell=1}^N H_{\rm b}(\rho_\ell)
$$
in which the limit exists since $0\leq H_{\rm b}(\rho_\ell)\leq 1$. This quantity can be interpreted as the limit average entropy of the cascade of $N$ BSCs and (\ref{eq:limit2}) can be interpreted as the asymptotic entropy rate of the correlated sources, namely the limit average source entropy. Equation (\ref{eq:limit2}) reduces, for the special case $\rho_{\ell}=\rho$, $\ell=1,2,\ldots,N$, to
$$
\overline{H}_{\rm b} = \lim_{N\rightarrow+\infty}\frac{1}{N}\sum_{\ell=1}^N H_{\rm b}(\rho) = H_{\rm b}(\rho) .
$$
This result can be also obtained by observing that $X_1^N$ is a stationary binary random process, whose entropy rate is well-known~\cite[Ch.~4]{CoTh06}. This limit for the special case of constant $\rho$ has been also derived in~\cite{AbFeMaFrRa14}.

Consider now the joint entropy of the serial correlated source model in Fig.~\ref{img:correlation}~(b). Using the chain rule for entropy, one can easily compute the joint entropy as
\begin{eqnarray}
H(X_1^N) &=& H(X_1) + \sum_{\ell=2}^N H(X_\ell|X_1^{\ell-1}) \nonumber \\
&=& H(X_1) + \sum_{\ell=2}^N H(X_\ell|X_{\ell-1}) \nonumber \\
&=& 1+\sum_{\ell=2}^N H_{\rm b}(\rho_\ell) . \label{eq:serial_entropy}
\end{eqnarray}
Analyzing the limit for large number of sources, one obtains
\begin{equation} \label{eq:limit3}
\lim_{N\rightarrow+\infty}\frac{H(X_1^N)}{N} = \overline{H}_{\rm b}
\end{equation}
and in the special case of $\rho_\ell=\rho$, $\ell=2,3,\ldots,N$:
$$
\overline{H}_{\rm b} = \lim_{N\rightarrow+\infty}\frac{1}{N}\left[1+\mathlarger{\sum}\limits_{\ell=2}^{N} H_{\rm b}(\rho_\ell)\right] = H_{\rm b}(\rho) .
$$
Note that the parallel and serial models have equal asymptotic source entropy rate.

In the general mixed case of Fig.~\ref{img:correlation}~(c), it can be shown (see Appendix~\ref{app:entropy_mixed} for the proof) that
\begin{equation} \label{eq:limit22}
\lim_{N\rightarrow+\infty}\frac{H(X_{11}^N,\ldots, X_{1M}^N)}{MN} = \overline{\overline{H}}_{\rm b}
\end{equation}
where
$$
\overline{\overline{H}}_{\rm b} \triangleq \lim_{N\rightarrow+\infty}\frac{1}{MN}\sum_{j=1}^M \sum_{\ell=1}^N H_{\rm b}(\rho_{\ell j}) .
$$
We remark that the limit exists since each term in the summation is limited to the interval $[0,1]$. Equation (\ref{eq:limit22}) reduces, for the special case of $\rho_{\ell j}=\rho$, $\ell=1,2,\ldots,N$ and $j=1,2,\ldots,M$, to
$$
\overline{\overline{H}}_{\rm b} = \lim_{N\rightarrow+\infty}\frac{1}{MN}\sum_{j=1}^M \sum_{\ell=1}^N H_{\rm b}(\rho) = H_{\rm b}(\rho) .
$$
Note that similar considerations can be also carried out if $N$ is kept fixed and $M$ grows to infinity or if both $N$ and $M$ become arbitrarily large.

We finally analyze the entropy rate of the source correlation model given by a set of linear equations in (\ref{eq:corr_model2}). Using (\ref{eq:identity}) and the assumption of invertibility of $\pmb{A}$, it can be easily shown that~\cite{CoTh06}
$$
H(X_1^N) = H(Z_1^N) = \sum_{\ell=1}^N H_{\rm b}(\rho_\ell) .
$$
Therefore:
$$
\lim_{N\rightarrow+\infty}\frac{H(X_1^N)}{N} = \overline{H}_{\rm b}
$$
which reduces, for the special case of $\rho_{\ell}=\rho$, $\ell=1,2,\ldots,N$, to
$$
\overline{H}_{\rm b} = \lim_{N\rightarrow+\infty}\frac{1}{N}\sum_{\ell=1}^N H_{\rm b}(\rho) = H_{\rm b}(\rho) .
$$
This shows that the source entropy rate is asymptotically the same for the linear equation-based correlation model as well, hence for all the considered correlation models.

The convergence of the entropy rates of the considered models can be analyzed in terms of the difference between the average source entropy for finite $N$ and the asymptotic value. In particular, in the parallel and serial models we define
$$
\epsilon \triangleq \dfrac{H(X_1^N)}{N} - \overline{H}_{\rm b} .
$$
In the mixed model, instead, this difference is also function of $M$ and can be defined as
$$
\epsilon \triangleq \dfrac{H(X_{11}^N,\ldots, X_{1M}^N)}{MN} - \overline{\overline{H}}_{\rm b} .
$$

\begin{figure}
\centering
\includegraphics[width=0.48\textwidth]{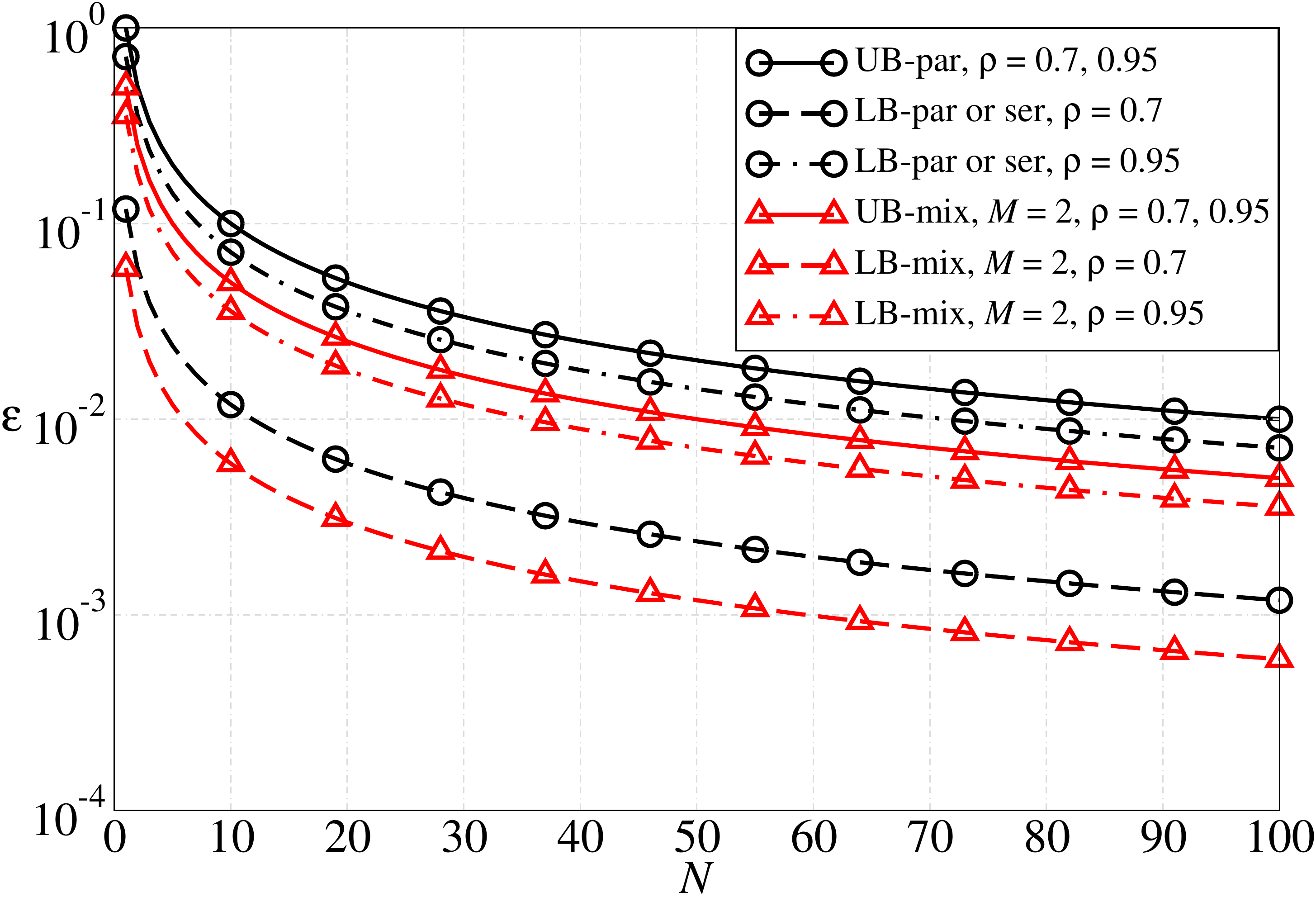}
\caption{$\epsilon$, as a function of $N$, for parallel (par), serial (ser), and mixed (mix) ($M=2$) models and two values of $\rho$: 0.7 and 0.95.}
\label{img:entropy_rate}
\end{figure}

In Fig.~\ref{img:entropy_rate}, $\epsilon$ is shown, as a function of $N$, for parallel or serial models and two values of $\rho$ (assumed equal for all BSCs): 0.7 and 0.95.
Note that lower and upper bounds on $\epsilon$ can be obtained using the LB and UB on the joint entropy for the parallel model, respectively. In particular the LB for the parallel case and the exact value of $\epsilon$ for the serial model are identical, due to the fact that the right-hand side of (\ref{eq:b}) and (\ref{eq:serial_entropy}) coincide for any $N$ in the special case of constant $\rho$. Moreover, the UB on $\epsilon$ for the parallel model is the same, regardless of the value of $\rho$, and the curves overlap for both values of $\rho$. In both cases, $\epsilon$ converges as $1/N$ and the same convergence can be obtained for the model given by a set of linear equations when $\pmb{A}$ is such that this model is equivalent to the serial one. Furthermore, the tightness of the bounds increases with the number of correlated sources, since UB and LB become closer to each other. However, the convergence of the LB degrades with increasing values of $\rho$. Moreover, the mixed model has the same trend for 
$\epsilon$, since it decreases with $1/(MN)$ for any fixed value of $M$.

In Fig.~\ref{img:entropy_rateM}, $\epsilon$ is shown, as a function of $M$, for the mixed model, two values of $\rho$ (namely, 0.7 and 0.95), and two values of $N$ (namely, 10 and 50). Note that similar considerations as those relative to Fig.~\ref{img:entropy_rate} hold in this case for fixed values of $N$ and letting $M$ go to infinity.

\section{Transmission of Correlated Symbols in Orthogonal Multiple Access Schemes\label{sec:achievable2}}
The computation of the source entropy rate discussed in Section~\ref{sec:achievable} plays an important role in determining the asymptotic achievable region of orthogonal multiple access schemes with correlated sources. In such schemes, each node independently encodes, by a joint source-channel code with rate $r$, the source symbols and transmits them through an orthogonal multiple access channel. At the receiver side, data are decoded by properly taking into account the source correlation to improve the overall system performance.

Recent work in~\cite{AbFeMaFrRa14} has proposed a characterization of the achievable region of orthogonal multiple access schemes with correlated sources, based on the computation of joint and conditional entropies of the sources, for an arbitrary value of~$N$. According to~\cite{AbFeMaFrRa14}, the achievable region, in the space of individual channel capacity values $\{\lambda_{\ell}\}_{\ell=1}^N$, is specified by the intersection of the following inequalities:
\begin{equation} \label{eq5NN}
\sum_{\ell\in\mathcal{S}} \lambda_{\ell} \ge r \, H(X(\mathcal{S})|X(\mathcal{S}_{\rm c}))
\end{equation}
for all $\mathcal{S}\subseteq\{1,2,\ldots,N\}$, in which $X(\mathcal{S})=\{X_i : i \in \mathcal{S}\}$ and $\mathcal{S}_{\rm c}$ denotes the complementary set of $\mathcal{S}$. Note that $H(X(\mathcal{S})|X(\mathcal{S}_{\rm c}))$ is the conditional entropy of the sources with index in $\mathcal{S}$ given the remaining ones. Two characteristic operational points, denoted as ``balanced'' and ``unbalanced,'' are of interest.

The \emph{balanced} characteristic point refers to the case where all source symbols are transmitted at a rate equal to the same single-channel capacity, i.e., $\lambda_{\rm bal}=\lambda_1 = \lambda_2 = \cdots = \lambda_N$. This common value, is equal to
\begin{equation} \label{eq:lbal}
\lambda_{\rm bal} \triangleq r\frac{H(X_1^N)}{N} .
\end{equation}

\begin{figure}
\centering
\includegraphics[width=0.48\textwidth]{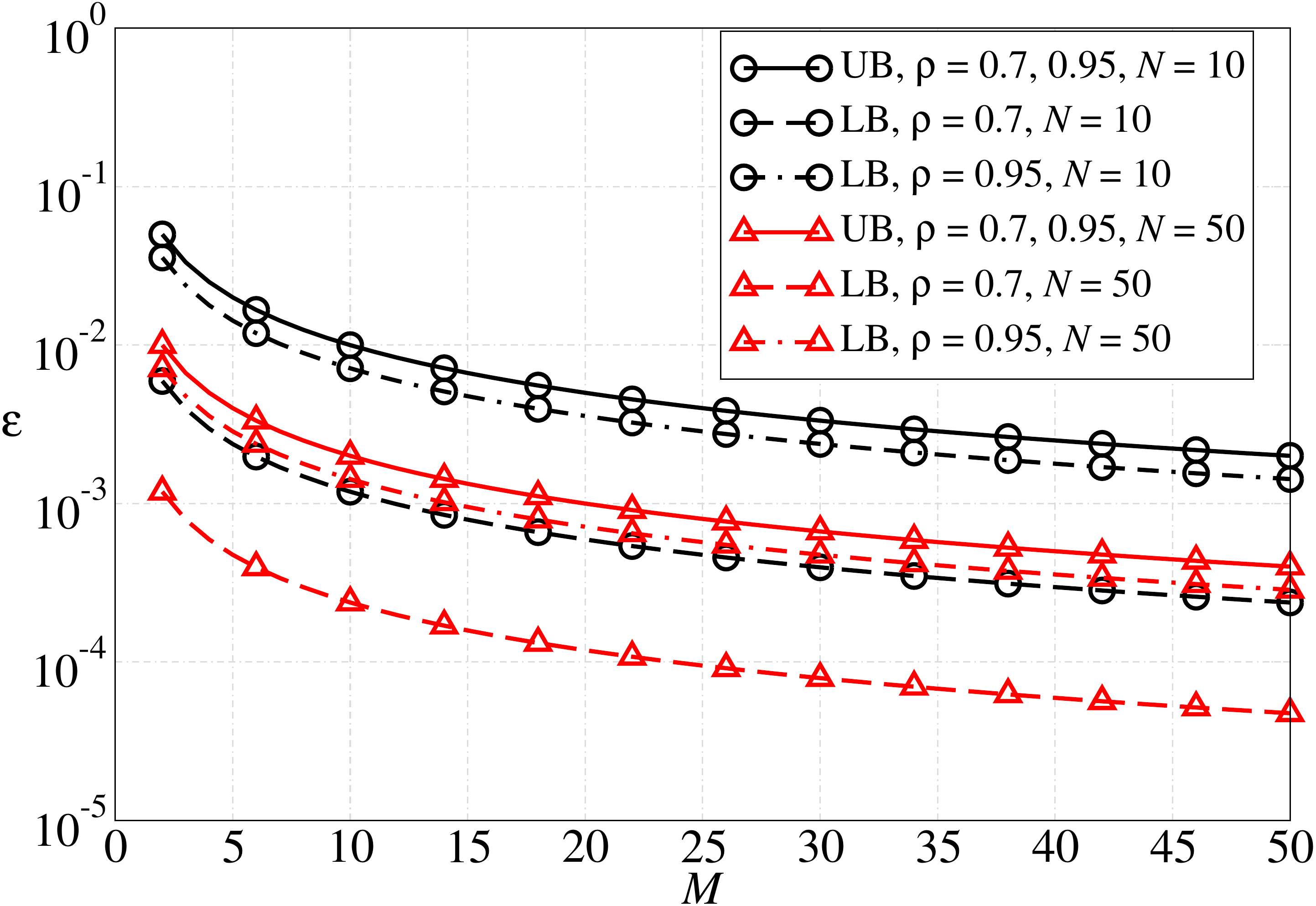}
\caption{$\epsilon$, as a function of $M$, for the mixed model, two values of $\rho$ (namely, 0.7 and 0.95), and two values of $N$ (namely, 10 and 50).}
\label{img:entropy_rateM}
\end{figure}

The \emph{unbalanced} case, instead, refers to the portion of the achievable region characterized as follows: $N-1$ sources, e.g., sources from 1 to $N-1$, are associated with sufficiently large values of $\lambda_i$, $i=1,2,\ldots,N-1$. In this case, $\lambda_{\rm unb}$ is the smallest value of $\lambda_N$ such that the operational point lies on the border of the achievable region and it is equal to
\begin{equation} \label{eq:lunb}
\lambda_{\rm unb} \triangleq r H(X_N|X_1^{N-1}) .
\end{equation}

In~\cite{AbFeMaFrRa14}, it is shown that for the parallel model in Fig.~\ref{img:correlation}~(a) and $\rho_\ell=\rho$, $\ell=1,2,\ldots,N$, the following facts hold
\begin{eqnarray}
&& \lambda_{\rm bal} \geq \lambda_{\rm unb} \qquad \forall N \nonumber \\
&&\lim_{N\rightarrow +\infty} \lambda_{\rm unb} = \lim_{N\rightarrow +\infty} \lambda_{\rm bal} \triangleq \lambda_{\rm lim} = r H_{\rm b}(\rho) . \label{eq:inf_rate}
\end{eqnarray}
This characterization of the achievable multiple access region is simple but effective. In particular, (\ref{eq:inf_rate}) tells us that, when $N$ increases, the achievable region tends to a hyperoctant and the system operational points become equal. Therefore, one can devise joint source-channel coded schemes for any one of these operational points, since they guarantee the same achievable rate of other operational points for a sufficiently large number of sources. Hence, code design can be based on the most convenient operational point, e.g., the one which guarantees less complexity.

A natural question, not discussed in~\cite{AbFeMaFrRa14}, is the following: are these results dependent on the considered correlation model? In general, the entropy of the binary sources may depend on the model considered for correlation. However, we have shown in Section~\ref{sec:achievable} that the asymptotic entropy rate is the same for all the considered models. We can, therefore, conclude that the achievable region of orthogonal multiple access schemes is asymptotically invariant to the considered correlation model as the number of sources increases. Moreover, from Figs.~\ref{img:entropy_rate} and~\ref{img:entropy_rateM} it can be observed that the convergence trend is the same for all considered models. Hence, it may be expected that joint source-channel coded schemes devised for one of these models (see, e.g., the turbo codes considered in~\cite{AbFeMaFrRa14} for the parallel model) have similar asymptotic performance for the other models as the number of sources increases.

\section{Conclusions\label{sec:Conclusions}}
In this paper, we have discussed and analyzed a few binary source correlation models. In particular, we have statistically characterized these models, in terms of joint PMF, covariance, and joint entropy of the source correlated sequence. Moreover, we have computed the asymptotic entropy rate for large number of sources, showing that a common value is obtained for all considered models. This result generalizes~\cite{AbFeMaFrRa14}, because the asymptotic achievable region of orthogonal multiple access schemes for large numbers of sources is shown to be invariant to the considered correlation model. Therefore, one can conjecture that joint source-channel coded schemes, which exploit the correlation at the receiver, may be expected to have similar asymptotic behavior regardless of the model.

\appendices

\section{Existence of the Inverse Matrix in the Linear Correlation Model (\ref{eq:corr_model2})} \label{app:invertible}
We now present a few special cases of matrix $\pmb{A}$ in~(\ref{eq:corr_model2}) where the inverse matrix exists. In particular, we consider two main classes: (i) matrices associated with the recursive model in (\ref{eq:literature_model}) and (ii) circulant matrices. To show the invertibility of these matrices, we can prove that their determinants are non-zero in the binary field.

The coefficients of $\pmb{A}$ for the recursive model in (\ref{eq:literature_model}) are the following:
$$
A_{\ell k} = \left\{
\begin{array}{lll}
1 & & \textrm{for $k=\ell$}\\[2mm]
\textrm{$0$ or $1$} & & \textrm{for $k=\ell-\min\{D,\ell-1\},\ldots,\ell-1$}\\[2mm]
0 & & \textrm{otherwise}
\end{array}
\right.
$$
for $\ell,k=1,2,\ldots,N$. Therefore, the recursive model is characterized by a Toeplitz matrix $\pmb{A}$. It is known that, out of all the size-$N$ Toeplitz matrices over a finite field of $q$ elements, a fraction $1-1/q$ (i.e., $1/2$ in our binary case) is non-singular~\cite{KaLo96}. The considered matrix associated with the recursive model (\ref{eq:literature_model}) is also lower triangular with equal element on the main diagonal, since $A_{\ell k}=0$ for $\ell<k$. Therefore, the determinant can be written as~\cite{HoJo02}:
$$
\det{\pmb{A}} = \prod_{\ell=1}^N A_{\ell \ell} = 1 .
$$
This proves that the inverse matrix exists for this case.

Consider now circulant matrices and restrict to those matrices with coefficients, for $\ell=1,2,\ldots,N$, of the form:
$$
A_{\ell k} = \left\{
\begin{array}{lll}
1 & & \textrm{for $k=\ell,\ell+1,\ldots,\ell+d-1 \mod N$}\\[2mm]
0 & & \textrm{otherwise}
\end{array}
\right.
$$
where $d$ is such that $d$ consecutive matrix coefficients are equal to 1 and the remaining are equal to zero. The $\mod N$ operation is needed to perform the circular shift of the rows. The determinant is known if $d$ is a prime~\cite{HoJo02}:
$$
\det{\pmb{A}} = \left\{
\begin{array}{lll}
0 & & \textrm{if $d\mid N$}\\[2mm]
d \mod 2 & & \textrm{otherwise}
\end{array}
\right.
$$
where the notation $d\mid N$ means that $d$ divides $N$. Therefore, circulant matrices admit an inverse if $d$ is an odd prime (i.e., $d\neq2$) and does not divide $N$.

\section{Asymptotic Entropy Rate of the Mixed Correlation Model} \label{app:entropy_mixed}
Let us consider the joint entropy of $X_{1j}^N$ and $B$:
\begin{equation} \label{eq:proof33}
H(B,X_{11}^N,\ldots, X_{1M}^N) = H(B) + H(X_{11}^N,\ldots, X_{1M}^N|B) .
\end{equation}
Given $B$, the set $X_{1j}^N$ is independent of $X_{1k}^N $ for $j\neq k$ and, therefore,
$$
H(X_{11}^N,\ldots, X_{1M}^N|B) = \sum_{j=1}^M H(X_{1j}^N|B) = \sum_{j=1}^M \sum_{\ell=1}^N H_{\rm b}(\rho_{\ell j}) .
$$
Noting again that $B$ is a uniformly distributed binary random variable with $H(B)=H_{\rm b}(0.5)=1$, one obtains:
$$
H(B,X_{11}^N,\ldots, X_{1M}^N) = 1 +  \sum_{j=1}^M \sum_{\ell=1}^N H_{\rm b}(\rho_{\ell j}) .
$$
As in (\ref{eq:proof33}), it is also possible to write
\begin{eqnarray*}
1+ \sum_{j=1}^M \sum_{\ell=1}^N H_{\rm b}(\rho_{\ell j}) &=& H(B,X_{11}^N,\ldots, X_{1M}^N) \\
&&\hspace*{-25mm}= H(X_{11}^N,\ldots, X_{1M}^N) + H(B|X_{11}^N,\ldots, X_{1M}^N)
\end{eqnarray*}
and, therefore,
\begin{eqnarray} 
H(X_{11}^N,\ldots, X_{1M}^N) &=& 1 + \sum_{j=1}^M \sum_{\ell=1}^N H_{\rm b}(\rho_{\ell j}) \nonumber \\
&&- H(B|X_{11}^N,\ldots, X_{1M}^N) \label{eq:hn_th33} .
\end{eqnarray}
Since by definition of entropy the last term is non negative, the following UB results
\begin{equation} \label{eq:aa}
H(X_{11}^N,\ldots, X_{1M}^N) \leq 1+ \sum_{j=1}^M \sum_{\ell=1}^N H_{\rm b}(\rho_{\ell j})
\end{equation}
Moreover, since conditioning reduces entropy~\cite{CoTh06}, it also follows that
$$
H(B|X_{11}^N,\ldots, X_{1M}^N) \leq H(B|X_{11}) = H_{\rm b}(\rho_{11}) .
$$
Using this in (\ref{eq:hn_th33}), one obtains the LB:
\begin{equation}
H(X_{11}^N,\ldots, X_{1M}^N) \geq 1 + \sum_{j=1}^M \sum_{\ell=1}^N H_{\rm b}(\rho_{\ell j}) - H_{\rm b}(\rho_{11}) . \label{eq:bb}
\end{equation}
Combining (\ref{eq:aa}) and (\ref{eq:bb}), one obtains the limit in (\ref{eq:limit22}).

\section*{Acknowledgments}
The authors would like to thank Andrea Abrardo for useful discussions.

\bibliographystyle{IEEEtran}
\bibliography{references}

\end{document}